\documentclass[12pt]{article}
\usepackage{amsmath,amssymb,amsfonts,amsthm,pifont}
\usepackage{eucal} 
\usepackage[dvips]{graphics}
\usepackage{pstricks,pst-plot,pst-node}
\newpsobject{showgrid}{psgrid}{subgriddiv=1,griddots=10,gridlabels=0pt}
\input epsf.tex

\newcommand{\C}{\mathbb{C}}
\newcommand{\Z}{\mathbb{Z}}
\newcommand{\R}{\mathbb{R}}
\newcommand{\cO}{\mathcal{O}}
\newcommand{\Pl}{\mathbf{P}^1}

\newcommand{\Ps}{\mathbf{P}^3}
\newcommand{\bP}{\mathbf{P}}
\newcommand{\cI}{\mathcal{I}}
\newcommand{\bT}{\mathbf{T}}
\newcommand{\Mb}{\overline{\mathcal{M}}}
\newcommand{\lan}{\left\langle}
\newcommand{\ran}{\right\rangle}
\newcommand{\bw}{\mathsf{w}}
\newcommand{\bW}{\mathsf{W}}
\newcommand{\cb}{\textup{\ding{114}}}
\newcommand{\bV}{\mathsf{V}}
\newcommand{\bp}{\mathbf{p}}

\DeclareMathOperator{\vdim}{vir\,\,dim}
\DeclareMathOperator{\codim}{codim}
\DeclareMathOperator{\Hilb}{Hilb}
\DeclareMathOperator{\Ext}{Ext}
\DeclareMathOperator{\ev}{ev}
\DeclareMathOperator{\ch}{ch}

\newtheorem{Conjecture}{Conjecture}

\begin{document}
\title{Random surfaces enumerating algebraic curves}
\author{Andrei Okounkov
\thanks{Partially supported by NSF and Packard Foundation}}
\date{} \maketitle

\section{Overview}

The discovery that a relation exists between the two topics 
in the title was made by physicists who viewed 
them as two approaches
to Feynman integral over all surfaces in string theory: one 
via direct discretization, the other through topological 
methods. A famous example 
is the celebrated conjecture by Witten connecting 
combinatorial tessellations of surfaces (conveniently 
enumerated by random matrix integrals) with intersection 
theory on the moduli spaces of curves, see \cite{W}.
Several mathematical proofs of this 
conjecture are now available \cite{Kon1,OP1,Mir}, but the 
exact mathematical match between the two theories
remains miraculous. 

The goal of this lecture is to describe an a 
priori different connection between enumeration
of algebraic curves and random surfaces. The 
underlying mathematical conjectures relating
Gromov-Witten and Donaldson-Thomas theory of 
a complex projective threefold $X$ were made in \cite{mnop}.  
Related physical proposal, first 
made in \cite{ORV} and developed in \cite{INOV}, played
an important role in development of these ideas.
A link to matrix integrals will be 
briefly explained at the end of the lecture. 

An occasion like this calls for 
a review, but instead
I chose to present
views that are largely conjectural, definitely 
not in their final form, but appealing 
and with large unifying power. These ideas were 
developed in collaboration with A.~Iqbal, D.~Maulik, 
N.~Nekrasov, R.~Pandharipande, N.~Reshetikhin, and 
C.~Vafa. I would like to thank the organizers for 
the opportunity to present them here and my coauthors 
for the joy of joint work.

\section{Enumerative geometry of curves}

\begin{figure}[hbtp]
  \begin{center}
   \scalebox{0.4}{\includegraphics{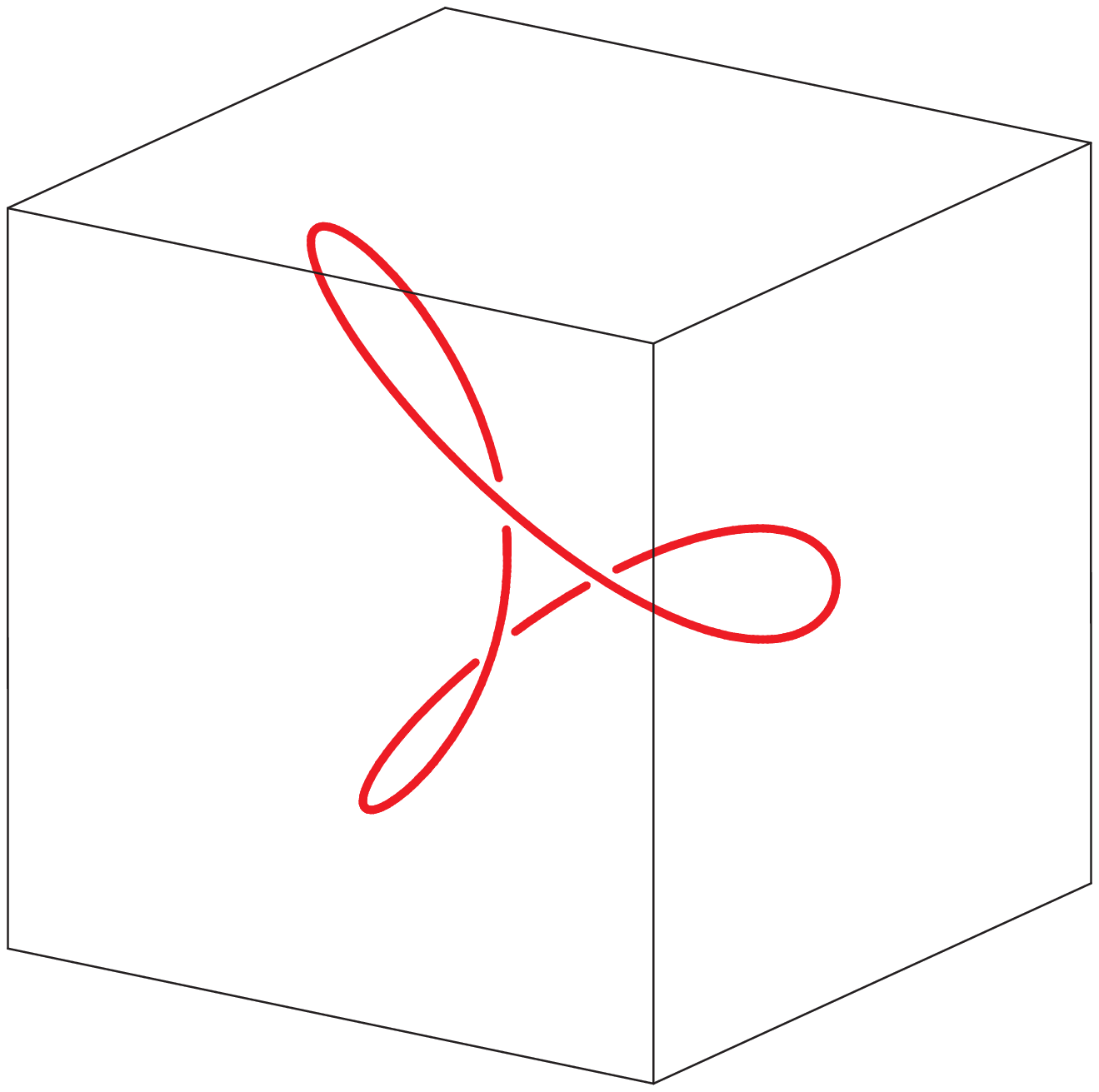}}     
    \caption{A degree 4 rational curve in $\R\Ps$}\label{f1}
     \end{center}
\end{figure}

Let $X$ be a smooth complex projective threefold
such as e.g.\ the projective space $\Ps$. We 
are interested in algebraic curves $C$ in $X$. 
For example, (the real locus of) a degree 4 
genus 0 curve in $\Ps$ may look like the one 
plotted in Figure \ref{f1}

Specifically, we are interested in enumerative 
geometry of curves in $X$. For example, we would like
to know how many curves
of given degree and genus meet given 
subvarieties of $X$, assuming we expect this number 
to be finite.

\subsection{Parametrized curves and stable maps}

\subsubsection{}

A rational curve $C$ in $X=\Ps$ like the one in Figure \ref{f1} 
is the image of the Riemann 
sphere $\Pl$ under a map
\begin{equation}
  \label{mapP}
  \Pl \in z \mapsto f(z)=\left[f_0(z) : f_1(z) : f_2(z) : f_3(z)\right]
\end{equation}
given in homogeneous coordinates 
by polynomials $f_i$ of degree $d$. Modulo reparameterization 
of $\Pl$, this leaves $4d$ complex parameters for $C$. 

To pass through a point in a threefold is a codimension 2  
condition on $C$. We, therefore, expect that finitely many
degree $d$ rational curves will meet $2d$ points 
in general position.  For example, there is obviously a unique 
line through two points. Similarly, since any conic 
lies in a plane, there will be none such 
passing through 4 generic points. In general, the number of 
degree $d=1,2,\dots$ rational curves through $2d$ general points
of $\Ps$ equals 
$$
1,0,1,4,105,2576,122129,\dots\,, 
$$
see for example \cite{CK,fp} on how to do such computations.
 
An important ingredient 
is a compactification of the space of maps \eqref{mapP} to 
the \emph{moduli space of stable maps}, introduced by 
Kontsevich. The domain of a stable map need not be 
irreducible, it may sprout off additional $\Pl$'s like 
in the case of a smooth conic degenerating to a union of 
two lines.

\subsubsection{}

In general, the moduli spaces $\Mb_{g,n}(X,\beta)$ of 
pointed stable maps to $X$ (where $X$ may be of any 
dimension) consist of data 
$$
(C,p_1,\dots,p_n,f)
$$
where $C$ is a complete curve of arithmetic genus $g$ with at 
worst nodal singularities, $p_1,\dots,p_n$ are smooth 
marked points of $C$, and $f:C\to X$ is an algebraic map 
of given degree 
$$
\beta = f_*([C])\in H_2(X)\,.
$$ 
Two such 
objects are identified if they differ by a reparameterization 
of the domain. One further requirement is that the 
group of automorphisms (that is, self-isomorphisms) should
be finite; this is the stability condition.

\subsubsection{}

The space $\Mb_{g,n}(X,\beta)$ carries a canonical 
\emph{virtual fundamental class} \cite{Beh,BehFan,LiTian} of dimension 
\begin{equation}
  \label{vdGW}
  \vdim \Mb_{g,n}(X,\beta) = 
- \beta \cdot  K_X + (g-1)(3-\dim X) + n \,, 
\end{equation}
where $K_X$ is the canonical class of $X$. 
The Gromov-Witten 
invariants of $X$ are defined as intersections of 
cohomology classes on $\Mb_{g,n}(X,\beta)$ defined
by conditions we impose on $f$ (e.g.\  by constraining 
the images $f(p_i)$ of the marked points) against the virtual fundamental 
class. In exceptionally good cases, for example when
$X=\Ps$ and $g=0$, the virtual fundamental class is 
the usual fundamental class.

Even for $X=\Ps$, the situation with higher genus curves 
is considerably more involved, both in foundational
aspects as well as in combinatorial complexity. 
It is, therefore, remarkable that conjectural 
correspondence with Donaldson-Thomas theory, to be 
described momentarily,  gives all-genera
fixed-degree 
answers with finite amount of computation.

\subsection{Equations of curves and Hilbert scheme}

Instead of giving a  parameterization, one can describe 
algebraic curves $C\subset X$ by their equations.

\subsubsection{}
Concretely, if $X\subset\bP^N$ for some $N$ and 
$[x_0:x_1:\dots:x_N]$ are homogeneous coordinates on 
$\bP^N$ then homogeneous 
polynomials $f$ 
vanishing on $C$ form a graded ideal 
$$
I(C) \subset \C[x_0,\dots,x_N] \,,
$$
containing the ideal $I(X)$ of $X$. This ideal is 
what replaces parametrization of $C$ in the world of 
equations. For example, the curve in Figure \ref{f1} is 
cut out (that is, its ideal is generated) 
by one quadratic and 3 cubic equations.

\subsubsection{}

Let 
$
I(C)_k \subset \C[x_0,\dots,x_N]_k
$
denote subspaces formed by polynomials of degree $k$. 
The codimension of $I(C)_k$ is the number of linearly 
independent degree $k$ polynomials on $C$. 
By Hilbert's theorem, 
\begin{equation}
  \label{HilbF}
  \codim I(C)_k = (\beta \cdot h) \, k + \chi(\cO_C) \,, \quad k\gg 0 \,,
\end{equation}
where $\beta\in H_2(X)$ is the class of $C$ and $h$ is the hyperplane 
class induced from the ambient $\bP^N$. The number 
$$
\chi(\cO_C)=\dim H^0(C,\cO_C) - \dim H^1(C,\cO_C)
$$ 
is the 
holomorphic Euler characteristic of $C$. By definition, 
$g=1-\chi$ is the arithmetic genus of $C$. 

It is easy to see that $C$ is 
uniquely determined by any $I(C)_k$ provided $k\gg 0$. 
A natural parameter space for ideals $I$ with given 
Hilbert function \eqref{HilbF} is the Hilbert
scheme $\Hilb(X;\beta,\chi)$ constructed by 
Grothendieck. It is defined by certain 
natural equations in the Grassmannian
of all possible linear subspaces 
$I_k\subset \C[x_0,\dots,x_N]_k$ 
of given codimension \eqref{HilbF}.

\subsubsection{}

While $\Hilb(X;\beta,\chi)$ and  $\Mb_{g,n}(X,\beta)$ 
play the same role of a compact parameter space in 
the world of equations and parameterizations, 
respectively, it should be stressed that there is no 
direct geometric relation between the two. This is 
most apparent in the case $\beta=0$. In degree $0$ 
case, the stable map moduli spaces become 
essentially Deligne-Mumford spaces of stable curves
--- very nice and well-understood varieties. 
The Hilbert scheme of points in a $3$-fold $X$, 
by contrast, seems very complicated. 
Even the number of its irreducible components, or
their dimensions, is not known.

\subsubsection{}

All of what we said so far about the Hilbert 
scheme applied very generally, in any dimension. 
The case of curves in a 3-fold, however, is 
special: in this case $\Hilb(X;\beta, \chi)$ 
carries a virtual fundamental class 
constructed by R.~Thomas \cite{Thom}. 
The technically important thing about 3-folds is
that Serre duality limits the number of interesting 
$\Ext^i$-group from an ideal sheaf to itself 
to just $i=1,2$. From \eqref{vdGW} we see that 
the case $\dim X=3$ is special for Gromov-Witten 
theory, too. In fact, we have
\begin{equation}
  \label{vdH}
  \vdim \Hilb(X;\beta,\chi) = \vdim \Mb_{g}(X,\beta) = - \beta \cdot  K_X \,.
\end{equation}
As we will see in the next section, it is very 
fortunate that this dimension depends only on $\beta$.

\subsection{Gromov-Witten and Donaldson-Thomas invariants}

Let $\beta\in H_2(X)$ be such that 
$-\beta \cdot  K_X \ge 0$. 
Let $\gamma_1,\dots,\gamma_n\in H_*(X)$ be a collection 
of cycles in $X$ such that 
$$
\sum (\codim \gamma_i -1)= -  \beta \cdot  K_X \,.
$$ 
By the dimension formula \eqref{vdH}, 
the virtual number of degree $\beta$ curves of some 
fixed genus meeting all of $\gamma_i$'s is expected
to be finite. 

The precise technical definition of this virtual 
number is different for stable maps and the Hilbert
scheme. 

\subsubsection{}

On the stable maps side, we can use marked 
points $p_i$ to say ``curve meets $\gamma_i$'' in 
the language of cohomology. Namely, 
imposing the condition $f(p_i)\in \gamma_i$ can be interpreted 
as pulling back the Poincar\'e dual class $\gamma_i^\vee$ 
via the evaluation map
\begin{equation}
 \ev_i : (C,p_1,\dots,p_n,f)\mapsto f(p_i)\,.
\end{equation}
The Gromov-Witten invariants are defined by 
\begin{equation}
  \label{GWi}
  \lan \gamma_1,\dots,\gamma_n \ran^{GW}_{\beta,g} = 
\int_{\left[\Mb_{g,n}(X,\beta)^\bullet\right]_{vir}}
\prod_{i=1}^n \ev_i^*\left(\gamma_i^\vee\right) \,. 
\end{equation}
The bullet here stands for  moduli space
with possibly disconnected domain and 
$[\quad]_{vir}$ is its virtual fundamental 
class. The disconnected theory contains, 
of course, the same information as the connected one, but has
slightly better formal properties. Most importantly, since 
connected curves don't form a component of the Hilbert scheme, 
we prefer to work with possibly
disconnected curves on the Gromov-Witten side as well.

\subsubsection{}\label{s_unis}

On the Hilbert scheme side, instead of marked points, it 
is natural to use characteristic classes of the universal 
ideal sheaf 
$$
\cI \to \Hilb(X) \times X \,, 
$$
which has the property that for any 
point $I\in \Hilb(X)$, the restriction of $\cI$ to 
$I \times X\cong X$ is $I$ itself. We have $c_1(\cI)=0$ and 
$$
c_2(\cI)\in H^2\left(\Hilb(X) \times X\right)
$$
can be interpreted as the class of locus
$$
\{(I,\text{point of the curve defined by $I$})\} \subset \Hilb(X) \times X \,.
$$
The class of curves $I\in\Hilb(X)$ meeting $\gamma\in H_*(X)$ can 
be described as the coefficient of $\gamma^\vee$ in the 
K\"unneth decomposition of $c_2(\cI)$. We denote this 
component by
$$
c_2(\gamma) \in H^{\codim \gamma -1}(\Hilb(X))
$$
and define 
\begin{equation}
  \label{DTi}
\lan \gamma_1,\dots,\gamma_n \ran^{DT}_{\beta,\chi} = 
\int_{\left[\Hilb(X;\beta,\chi)\right]_{vir}}
\prod_{i=1}^n  c_2(\gamma_i)\,. 
  \end{equation}
We call these numbers the Donaldson-Thomas invariants 
of $X$.

\subsection{Main conjecture}

\subsubsection{}

As already pointed out, there is no reason for the 
corresponding invariants \eqref{GWi}
and \eqref{DTi} to agree and, in fact, they don't. 
For one thing, the moduli spaces are empty and, hence, 
integrals vanish if $g,\chi\ll 0$, which goes
in the opposite directions via $\chi=1-g$. Also,  
the Donaldson-Thomas invariants are
integers while the Gromov-Witten invariants are typically 
fractions (because stable maps can have finite 
automorphisms). However, a conjecture proposed in \cite{mnop}
equates natural generating functions for the two kinds 
of invariants after a nontrivial change of variables.

\subsubsection{}

Concretely, set
$$
Z_{GW}(\gamma_1,\dots,\gamma_n;u)_\beta = \sum_{g} u^{2g-2} \, 
\lan \gamma_1,\dots,\gamma_n \ran^{GW}_{\beta,g}
$$
and define the reduced partition function by
$$
Z'_{GW}(\gamma;u)_\beta  = 
Z_{GW}(\gamma;u)_\beta \big/ Z_{GW}(\varnothing;u)_0 \,.
$$
This reduced partition function counts maps without 
collapsed connected components. The degree zero 
function $Z_{GW}(\varnothing;u)_0$ is known 
explicitly for any 3-fold $X$ by the results of \cite{FabPan}, 
see below. Define $Z_{DT}(\gamma;q)_\beta$
and its reduced version by the same formula, with 
$q^\chi$ replacing $u^{2g-2}$. 

\begin{Conjecture}\label{c_m} The reduced Donaldson-Thomas 
partition function $Z'_{DT}(\gamma;q)_\beta$ is 
a rational function of $q$. The change of variables
$$
\boxed{q=-e^{iu}}
$$
relates it to the Gromov-Witten partition functions
$$
(-iu)^{-\vdim} Z'_{GW}(\gamma;u)_\beta  = 
(-q)^{-\vdim/2} Z'_{DT}(\gamma;q)_\beta \,,
$$
where $\vdim=-\beta\cdot K_X$ is the virtual 
dimension.  
\end{Conjecture}

\subsubsection{}
Conjecture \ref{c_m} has been established 
when $X$ is either a local curve, that is, an 
arbitrary rank 2 bundle over a smooth curve \cite{GWDT} or 
the total space of canonical bundle over a 
smooth toric surface \cite{mnop,LLZ2}. 
In the local curve case, 
equivariant theory is needed \cite{jbrp}. In my
opinion, this provides substantial evidence
for the ``GW=DT''  
correspondence.

\subsubsection{}

Conjecture \ref{c_m} is actually a special 
case of more general conjectures proposed 
in \cite{mnop} that extend the GW/DT correspondence
to the relative context and descendent 
invariants. On the Gromov-Witten side, the 
descendent insertions are defined by  
$$
\tau_k(\gamma_i) = \ev_i^*(\gamma_i^\vee) \, \psi_i^{k}
\in H^{\codim \gamma_i + k}(\Mb_{g,n}(X;\beta))\,,
$$
where $\psi_i$ is the 1st Chern class of the 
line bundle $L_i$ over $\Mb_{g,n}(X;\beta)$ 
with fiber the cotangent line $T^*_{p_i}C$ to 
the curve $C$ at the marked point $p_i$.  
These should correspond to K\"unneth components
of characteristic classes of the universal 
sheaf $\cI$. For example, we conjecture that
$$
\tau_k(pt) \xrightarrow{\quad\textup{GW=DT}\quad}(-1)^{k+1} \ch_{k+2}(pt)\,,
$$
provided $\codim \gamma_i >0 $ for all 
other insertions. Here 
$$
\ch_{k+2}(\cI) \in H^{k+2}(\Hilb(X)\times X) 
$$ 
are the components of the Chern character of $\cI$ and 
$\ch_{k+2}(pt)$ are the coefficient of
$pt^\vee = 1\in H^*(X)$ in their K\"unneth 
decomposition.

\subsubsection{}

In the degree $0$ case, which is left out by Conjecture
\ref{c_m}, we expect the following simple answer 
which depends only on characteristic numbers of $X$. 
Denote the Chern classes of $TX$ by $c_i$ and 
let 
\begin{equation}
  \label{McM}
  M(q) = \prod_{n>0} (1-q^n)^{-n} 
\end{equation}
be the McMahon function.

\begin{Conjecture} \label{ZDT0}
$\displaystyle \quad 
Z_{DT}(X,q)_0 = M(-q)^{\int_X(c_3-c_1 c_2)}\,. 
$
\end{Conjecture}

This conjecture has been proven for a large class of 
3-folds including all toric ones \cite{mnop}. 

Comparing the asymptotic expansion 
\begin{equation}
  \label{McMa}
   \ln M(e^{-u}) \sim \sum_{g=0}^{\infty} 
\frac{\zeta(3-2g) \zeta(1-2g)}{(2g-2)!} \, u^{2g-2} \,, 
\quad u\to +0 \,. 
\end{equation}
in which the singular $g=1$ term is understood 
as the second term in 
$$
\frac{\zeta(3-2g) \zeta(1-2g)}{(2g-2)!} \, u^{2g-2} =
\frac{1}{24}\frac{1}{g-1} + 
\left(\frac1{12}\ln u + \zeta'(-1)\right) + O(g-1) \,, 
$$
to evaluation of $Z_{GW}(X,u)_0$ obtained
in \cite{FabPan}, we find 
$$
\ln Z_{DT}(X,-e^{iu})_0 \sim \dots + 2 
\ln Z_{GW}(X,u)_0 \,,
$$
where dots stand for singular or constant terms in 
the asymptotic expansion. 
There are some plausible explanations for the unexpected 
factor of $2$ in this formula, but none convincing enough to be 
presented here. 

McMahon's discovery was that the function $M(q)$ is
 the generating 
function for 3-dimensional partitions. We will see
momentarily how 3-dimensional partitions arise in 
Donaldson-Thomas theory.

\section{Random surfaces}

\subsection{Localization and dissolving crystals}

\subsubsection{}

For the rest of this lecture, we will assume that 
$X$ is a smooth toric $3$-fold, such as $\Ps$ or $(\Pl)^3$. 
By definition, this means, that the torus $\bT=(\C^*)^3$ 
acts on $X$ with an open orbit. Since anything that acts on $X$
naturally acts on both $\Mb_{g,n}(X;\beta)$ and $\Hilb(X;\beta,\chi)$, 
localization in $\bT$-equivariant cohomology \cite{AtBo} can 
be used to compute intersections on these moduli spaces, see 
\cite{ES,Kon2,GP}. 

Localization reduces intersection computations
to certain integrals over the loci of $\bT$-fixed points. 
On the Gromov-Witten side, these fixed loci are, 
essentially, moduli spaces of curves and the 
integrals in question are the so-called Hodge
integrals. While any fixed-genus Hodge integral can, 
in principle, be evaluated in finite time, 
a better structural understanding of the totality 
of these numbers remains an important challenge. 
By contrast, the $\bT$-fixed loci in the Hilbert 
scheme are isolated points. Together with the 
conjectural rationality of $Z'_{DT}$, this 
reduces, for fixed degree, the
all-genera answer to a finite sum. 

\subsubsection{}
It is the localization sum in the Donaldson-Thomas 
theory that can be interpreted as the partition 
function of a certain random surface ensemble. 
The link is provided by the combinatorial geometry 
of the $\bT$-fixed points in the Hilbert scheme,
which is standard and will be quickly reviewed 
now.

\subsubsection{}

As a warm-up, let us start with surfaces instead of $3$-folds and 
look at the Hilbert scheme $\Hilb(\C^2;d,n)$ formed
by ideals $I\subset \C[x,y]$ such that
\begin{equation}
  \label{dnH}
  \codim I_{\le k} = d k + n \,, \quad k \gg 0 \,, 
\end{equation}
where $I_{\le k}$ stands for subspace of polynomials
of degree $\le k$. The torus $(\C^*)^2$ acts on 
$\Hilb(\C^2;d,n)$ by rescaling $x$ and $y$. The 
monomials $x^i y^j$ are eigenvectors of the 
torus action with distinct eigenvalues. Any 
torus-fixed linear subspace $I\subset \C[x,y]$ 
is, therefore, spanned by monomials. Since
$I$ is also an ideal, together 
with any monomial $x^i y^j$ it contains all 
monomials $x^a y^b$ with $a\ge i$ and $b\ge j$.

\begin{figure}[hbtp]\psset{unit=1.1 cm}
  \begin{center}
    \begin{pspicture}(0,0)(10,7)
\newgray{lightgray}{.90}
\psframe[fillstyle=solid,linewidth=0,fillcolor=lightgray,
linecolor=white](1,0)(10,3)
\psframe[fillstyle=solid,linewidth=0,fillcolor=lightgray,
linecolor=white](3,3)(10,4)
\psframe[fillstyle=solid,linewidth=0,fillcolor=lightgray,
linecolor=white](4,4)(10,5)
\psframe[fillstyle=solid,linewidth=0,fillcolor=lightgray,
linecolor=white](6,5)(10,6)
\showgrid
\psline[linewidth=0.05](1,0)(1,3)(3,3)(3,4)(4,4)(4,5)(6,5)(6,6)(10,6)
\pscircle(1.5,2.5){0.5}
\pscircle(3.5,3.5){0.5}
\pscircle(4.5,4.5){0.5}
\pscircle(6.5,5.5){0.5}
\rput[c](0.5,6.5){$1$}
\rput[c](0.5,5.5){$y$}
\rput[c](0.5,4.5){$y^{2}$}
\rput[c](0.5,3.5){$y^{3}$}
\rput[c](0.5,2.5){$y^{4}$}
\rput[c](0.5,1.5){$y^{5}$}
\rput[c](0.5,0.5){$y^{6}$}
\rput[c](1.5,6.5){$x$}
\rput[c](1.5,5.5){$xy$}
\rput[c](1.5,4.5){$xy^{2}$}
\rput[c](1.5,3.5){$xy^{3}$}
\rput[c](1.5,2.5){$xy^{4}$}
\rput[c](1.5,1.5){$xy^{5}$}
\rput[c](1.5,0.5){$xy^{6}$}
\rput[c](2.5,6.5){$x^{2}$}
\rput[c](2.5,5.5){$x^{2}y$}
\rput[c](2.5,4.5){$x^{2}y^{2}$}
\rput[c](2.5,3.5){$x^{2}y^{3}$}
\rput[c](2.5,2.5){$x^{2}y^{4}$}
\rput[c](2.5,1.5){$x^{2}y^{5}$}
\rput[c](2.5,0.5){$x^{2}y^{6}$}
\rput[c](3.5,6.5){$x^{3}$}
\rput[c](3.5,5.5){$x^{3}y$}
\rput[c](3.5,4.5){$x^{3}y^{2}$}
\rput[c](3.5,3.5){$x^{3}y^{3}$}
\rput[c](3.5,2.5){$x^{3}y^{4}$}
\rput[c](3.5,1.5){$x^{3}y^{5}$}
\rput[c](3.5,0.5){$x^{3}y^{6}$}
\rput[c](4.5,6.5){$x^{4}$}
\rput[c](4.5,5.5){$x^{4}y$}
\rput[c](4.5,4.5){$x^{4}y^{2}$}
\rput[c](4.5,3.5){$x^{4}y^{3}$}
\rput[c](4.5,2.5){$x^{4}y^{4}$}
\rput[c](4.5,1.5){$x^{4}y^{5}$}
\rput[c](4.5,0.5){$x^{4}y^{6}$}
\rput[c](5.5,6.5){$x^{5}$}
\rput[c](5.5,5.5){$x^{5}y$}
\rput[c](5.5,4.5){$x^{5}y^{2}$}
\rput[c](5.5,3.5){$x^{5}y^{3}$}
\rput[c](5.5,2.5){$x^{5}y^{4}$}
\rput[c](5.5,1.5){$x^{5}y^{5}$}
\rput[c](5.5,0.5){$x^{5}y^{6}$}
\rput[c](6.5,6.5){$x^{6}$}
\rput[c](6.5,5.5){$x^{6}y$}
\rput[c](6.5,4.5){$x^{6}y^{2}$}
\rput[c](6.5,3.5){$x^{6}y^{3}$}
\rput[c](6.5,2.5){$x^{6}y^{4}$}
\rput[c](6.5,1.5){$x^{6}y^{5}$}
\rput[c](6.5,0.5){$x^{6}y^{6}$}
\rput[c](7.5,6.5){$x^{7}$}
\rput[c](7.5,5.5){$x^{7}y$}
\rput[c](7.5,4.5){$x^{7}y^{2}$}
\rput[c](7.5,3.5){$x^{7}y^{3}$}
\rput[c](7.5,2.5){$x^{7}y^{4}$}
\rput[c](7.5,1.5){$x^{7}y^{5}$}
\rput[c](7.5,0.5){$x^{7}y^{6}$}
\rput[c](8.5,6.5){$x^{8}$}
\rput[c](8.5,5.5){$x^{8}y$}
\rput[c](8.5,4.5){$x^{8}y^{2}$}
\rput[c](8.5,3.5){$x^{8}y^{3}$}
\rput[c](8.5,2.5){$x^{8}y^{4}$}
\rput[c](8.5,1.5){$x^{8}y^{5}$}
\rput[c](8.5,0.5){$x^{8}y^{6}$}
\rput[c](9.5,6.5){$x^{9}$}
\rput[c](9.5,5.5){$x^{9}y$}
\rput[c](9.5,4.5){$x^{9}y^{2}$}
\rput[c](9.5,3.5){$x^{9}y^{3}$}
\rput[c](9.5,2.5){$x^{9}y^{4}$}
\rput[c](9.5,1.5){$x^{9}y^{5}$}
\rput[c](9.5,0.5){$x^{9}y^{6}$}
\end{pspicture}
 \caption{A typical monomial ideal $I\subset \C[x,y]$.}
    \label{f2}
\end{center}
\end{figure}

See Figure \ref{f2} for an image of a typical torus-fixed ideal
$I$. Monomials in the ideal $I$ are
shaded gray; the generators of $I$ are circled. 
Monomials not in $I$ form a shape similar to the 
diagram of a partition, except that it has some infinite
rows and columns. The total width of these infinite
rows and columns ($2$, in this example) is the degree 
$d$ in \eqref{dnH}. The constant term $\chi$ ($=9$ here) 
can be interpreted as the ``renormalized area'' of this infinite 
diagram.

\subsubsection{}

\begin{figure}[hbtp]
  \begin{center}
   \scalebox{0.7}{\includegraphics{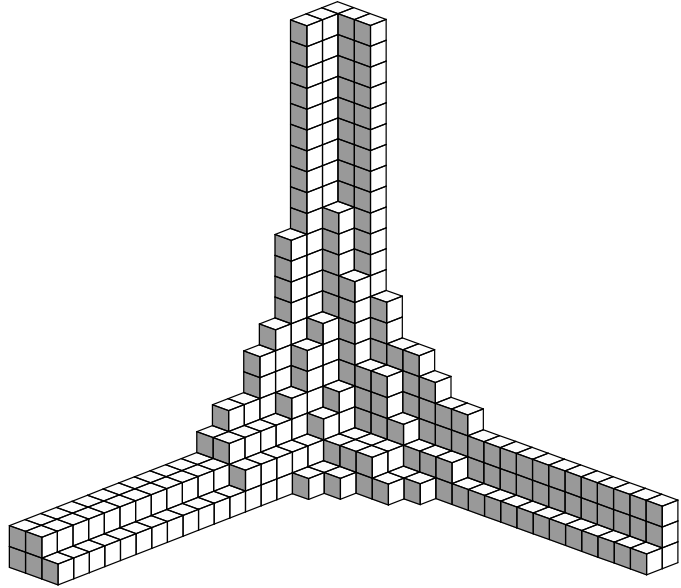}}     
    \caption{A monomial ideal in $\Hilb(C^3;d,\chi)$}\label{f3}
     \end{center}
\end{figure}

For $\Hilb(\C^3;d,\chi)$, the 
description of $\bT$-fixed points $I$ is similar, but now 
in terms of 3-dimensional partitions, with 
possibly infinite legs along the coordinate axes, 
see Figure \ref{f3}.  The 2D 
partitions $\lambda_1,\lambda_2,\lambda_3$, 
on which the infinite legs end, describe the 
nonreduced structure of $I$ along the coordinate axes.
The degree 
$$
d=|\lambda_1|+|\lambda_2|+|\lambda_3|
$$
is the total cross-section 
of the infinite legs; the number $\chi$ is
the renormalized volume of this 3D partition. 

\begin{figure}[hbtp]
  \begin{center}
   \scalebox{0.4}{\includegraphics{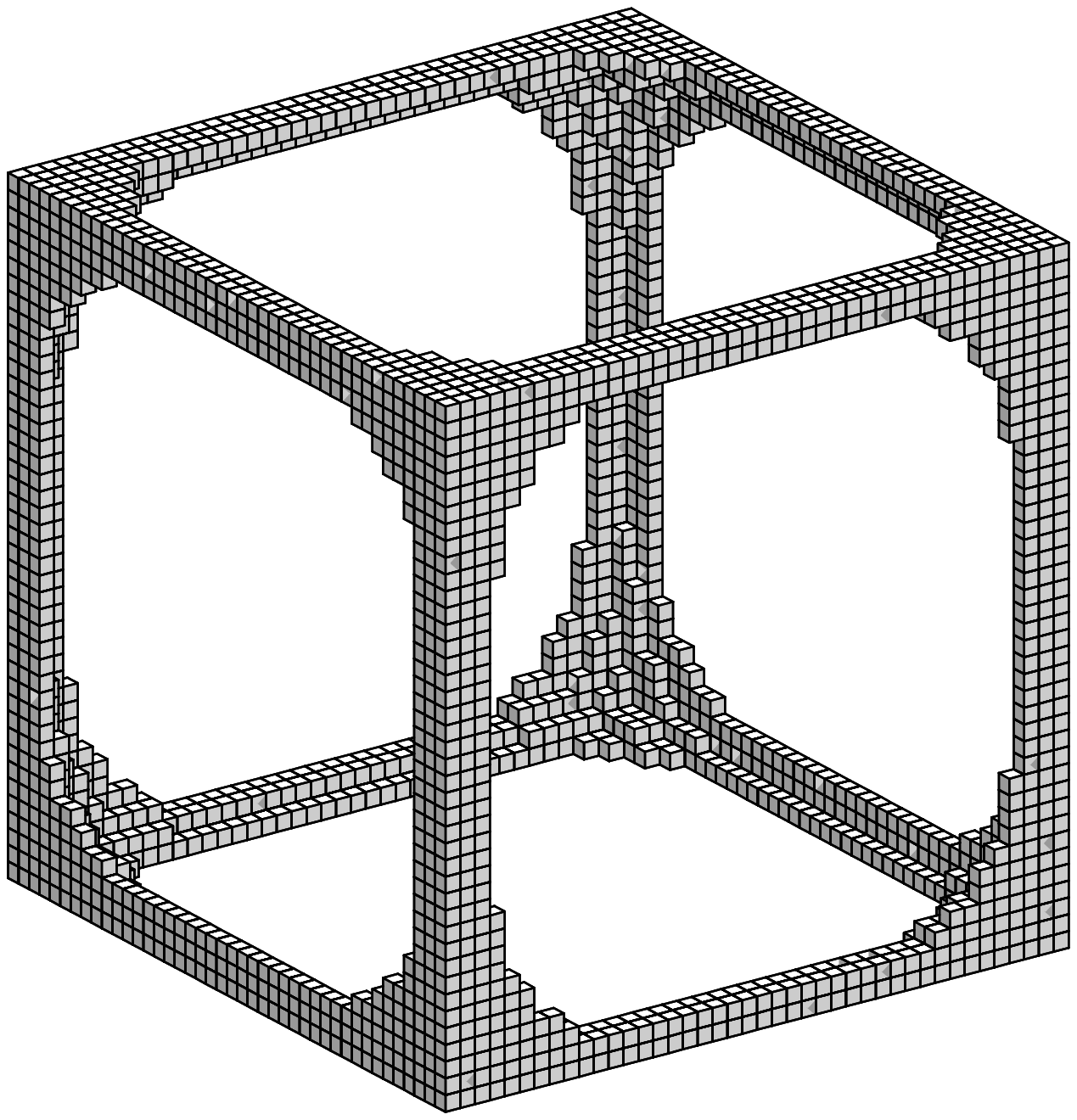}} \qquad 
  \scalebox{0.27}{\includegraphics{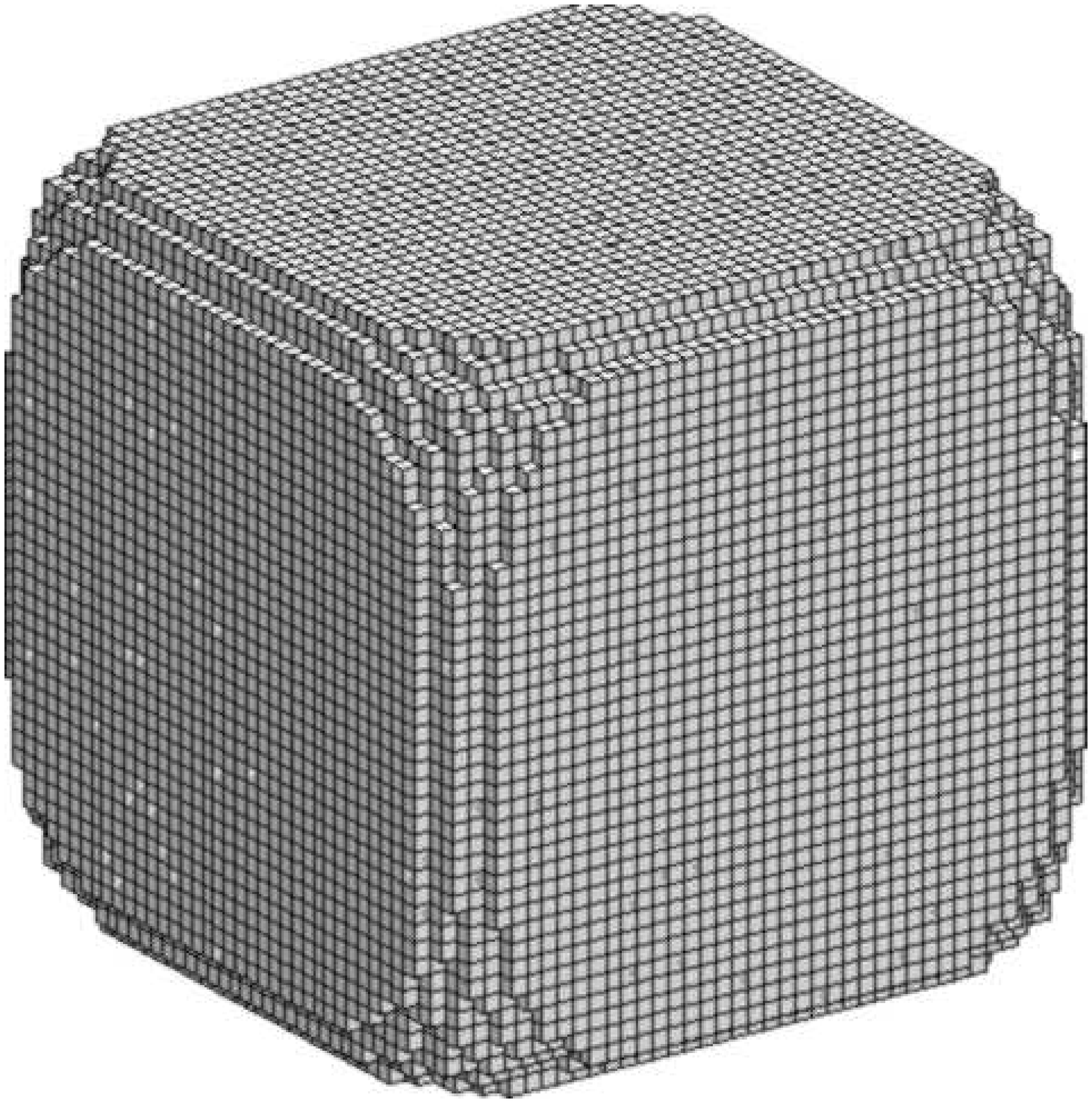}}
      \caption{A $\bT$-fixed point in $\Hilb((\Pl)^3;\beta,\chi)$}\label{f4}
     \end{center}
\end{figure}

A general projective toric $X$ corresponds to 
lattice polytope $\Delta_X$, with vertices corresponding to 
$\bT$-fixed points, edges --- to $\bT$-invariants $\Pl$'s et
cetera. For example, $(\Pl)^3$ and $\Ps$ corresponds to a cube
and simplex, respectively. To specify a $\bT$-fixed point in $\Hilb(X;\beta,\chi)$, 
we place a 3D partition at every vertex of $\Delta_X$. 
These 3D partition may have infinite legs along the 
edges of $\Delta_X$; we require that these legs glue 
in an obvious way, see Figure \ref{f4}, left half. We have 
$$
\beta = \sum_{\textup{edges $E$}} |\lambda_E| \, \left[E\right]\in H_2(X) \,,
$$
where $[E]$ is the class of the $\bT$-invariant $\Pl$ 
corresponding to the edge $E$ and $\lambda_E$ is 
cross-section profile along $E$. The number $\chi$ is 
the renormalized volume of this assembly of 3D partitions.

Note that 
the edge lengths do not have any intrinsic meaning 
in Figure \ref{f4}; formally, they have to be viewed
as infinitely long. It is an interesting problem to 
construct a generalization of Donaldson-Thomas theory 
in which the edge lengths will play a role. This 
should involve doubling of the degree parameters in 
the theory. 

The right half of Figure \ref{f4} shows the complement
of the 3D partition structure on the left. Note that 
it is highly reminiscent of a partially dissolved 
cubic crystal --- some atoms are missing from the corners
and along the edges. So, at least as far as the 
index set is concerned, the localization sum in Donaldson-Thomas
theory of $X$ has the shape of a partition function 
in a random surface model, the surface being the 
surface of the dissolving crystal. We now move on 
to the computation of localization weight.

\subsection{Equivariant vertex}

The weight of a $\bT$-fixed point $I\in\Hilb(X;\beta,\chi)$ 
in the virtual localization formula for Donaldson-Thomas
invariants was computed in \cite{mnop}. Here, for simplicity, 
we focus on the case $X=\C^3$ and $\beta=0$, that is, on the 
case of a single 3D partition without infinite legs. 
The general case is parallel.

\subsubsection{}

Let $I_\pi\in\Hilb(\C^3;0,\chi)$ be a monomial ideal 
corresponding to a 3D partition $\pi \subset \Z_{\ge 0}^3$. 
Let $C_\pi \subset \Z_{\ge 0}^3$ denote the complement 
of $\pi$; we view the elements of $C_\pi$ as the atoms 
that remain in the crystal. 

Let $z\in \C^* \subset \bT$ act on the coordinates in $\C^3$ by
$$
z \cdot (x_1,x_2,x_3) = (z^{t_1} x_1, z^{t_2} x_2, z^{t_3} x_3) \,.
$$ 
The localization weight $\bw(\pi)$ of $I_\pi$ 
will be a rational function of the parameters $t_i$. Let $T$ 
be the linear function taking value
$$
T(\cb)=t_1 a_1 + t_2 a_2 + t_3 a_3 \,.
$$
on a box
$\cb = (a_1,a_2,a_3) \in \Z_{\ge 0}^3$.
For a pair of boxes $\cb_1$ and $\cb_2$, we define
$$
U(\cb_1,\cb_2) = \frac{\delta T (\delta T+t_1+t_2)(\delta T+t_1+t_3)(\delta T+t_2+t_3)}
{(\delta T+t_1)(\delta T+t_2)(\delta T+t_3)(\delta T+t_1+t_2+t_3)} \,, 
$$
where 
$$
\delta T= T(\cb_1) - T(\cb_2) \,.
$$

Recall that $\chi$ is the number of missing atoms. 
We would have liked to define $\bw(\pi)$ by
\begin{equation}
  \label{w_prod}
  \bw(\pi)\textup{ ``$=$'' } (-q)^{\chi} \prod_{\cb_1,\cb_2 \in 
\textup{ crystal $C_\pi$}} U(\cb_1,\cb_2) \,,
\end{equation}
which has a standard grand-canonical Gibbs form
with $(-q)$ being the fugacity and 
$$
- \log U(\cb_1,\cb_2)\, U(\cb_2,\cb_1)
$$
being the (translation-invariant) interaction energy between 
the atoms in positions $\cb_1$ and $\cb_2$.

\subsubsection{}

Since the product \eqref{w_prod} is not even 
close to being well-defined or convergent, the following
regularization is required. Define
\begin{equation}
  \label{R_pi}
   R_\pi(z) = \textup{trace of $z$ acting on $I_\pi$}
 = \sum_{\cb\in C_\pi} z^{T(\cb)} \,,
\end{equation}
This can be viewed as a generating function of the set $C_\pi$.  
One checks that for any 3D partition $\pi$ 
\begin{equation}
  \label{Vpi}
  \bV_\pi(z) =- \frac{R_\pi(z) \, R_\pi(z^{-1})}{R_\varnothing(z^{-1})}+ R_\varnothing(z)
\end{equation}
is a Laurent polynomials in $z^{t_i}$, that is, it has the 
form 
$$
\bV_\pi(z) = \sum_{a\in \Z^3} v_\pi(a) \, z^{T(a)} \,, \quad v_\pi(a)\in \Z\,,
$$
where the sum is finite, that is, $v_\pi(a)=0$ for all but finitely 
many $a$.  We define the \emph{equivariant vertex measure} of 
a 3D partition $\pi$ by  
\begin{equation*}
  \label{w_ren}
   \bw(\pi)= q^{\chi} \prod_{a\in \Z} T(a)^{-v_\pi(a)} \,. 
\end{equation*}
Note that a naive expansion of the $R_\pi(z) \, R_\pi(z^{-1})$ product in 
\eqref{Vpi} leads to the infinite product in \eqref{w_prod}.

\subsubsection{}

It is a theorem from \cite{mnop} that the virtual fundamental class
of the Hilbert scheme restricts to the $\bT$-fixed point $I_\pi$ 
as follows: 
$$
q^\chi \left[\Hilb(\C^3;0,\chi)\right]_{vir}\bigg|_{I_\pi}  = \bw(\pi) \,. 
$$

\subsubsection{}

One special case worth noting is when 
\begin{equation}
  \label{CY}
   t_1 + t_2 + t_3 = 0 \,. 
\end{equation}
In this case 
$$
U(\cb_1,\cb_2) \, U(\cb_2,\cb_1) = 1 
$$
and the equivariant vertex measure $\bw$ becomes uniform on 
partitions of fixed size. Condition \eqref{CY} is the 
Calabi-Yau condition, it means restriction to 
the subtorus in $\bT$ preserving the holomorphic
$3$-form 
$$
\Omega=dx_1 \wedge dx_2 \wedge dx_3
$$ 
on $\C^3$. This explains why the McMahon function \eqref{McM}
appears in Donaldson-Thomas theory. 

For general $t_i$, the 
 analog of McMahon's identity is the 
following formula proven in \cite{mnop}
\begin{equation}
  \label{DTMc}
  \sum_\pi \bw(\pi)  = M(-q)^{-\tfrac{(t_1+t_2)(t_1+t_3)(t_2+t_3)}{t_1 t_2 t_3}} \,. 
\end{equation}
This formula implies Conjecture \ref{ZDT0} for any toric $3$-fold $X$.

\subsubsection{}
If $\pi$ has infinite legs, additional counterterms 
are needed in \eqref{Vpi} to make it finite and the 
measure $\bw(\pi)$ well-defined \cite{mnop}. The \emph{equivariant
vertex} is a function of 3 partitions $\lambda,\mu,\nu$ 
defined by 
\begin{equation}
  \label{bW}
  \bW(\lambda,\mu,\nu) = \sum_{\textup{$\pi$ ending on $\lambda,\mu,\nu$}}
\bw(\pi) \,.
\end{equation}
This function, which is the main building block in 
localization formula for Donaldson-Thomas invariants, 
is, in general, rather intricate. Conjecturally it is 
related to general triple Hodge integrals. In the 
Calabi-Yau case \eqref{CY} it specializes to 
the \emph{topological vertex} \cite{topver,ORV}, which has an expression 
in terms of Schur functions. The conjectural relation to 
Hodge integrals is proven in the one-leg case  \cite{GWDT}. 
In the much simpler Calabi-Yau case, it is known in 
the two-leg case, see \cite{LLZ2} and also \cite{LLZ1,OP5,LLLZ}.

\subsubsection{}

Conjecture \ref{c_m} relates the Donaldson-Thomas partition 
function $Z_{DT}$, which we just interpreted as the partition 
function of a certain dissolving crystal model, to the 
the Gromov-Witten partition function via the substitution
$$
- q = e^{iu} \,.
$$
This means that the asymptotic expansion of 
the free energy $\ln Z_{DT}$ in the 
thermodynamic limit
$$
-q= \textup{fugacity} \to 1
$$
gives a genus-by-genus count of connected curves in Gromov-Witten 
theory. Letting $q\to -1$ does corresponds to letting the energy cost 
of removal of an atom from the crystal go to zero. 
As a result, the expected number of removed atoms
\begin{equation}
\lan |\pi| \ran_{\bw} 
\, \overset{\textup{def}}{=}\,
 \frac{\sum \bw(\pi) \, |\pi|}{ \sum \bw(\pi)}  
\sim  \frac{(t_1+t_2)(t_1+t_3)(t_2+t_3)}{t_1 t_2 t_3} \, 
\frac{2\zeta(3)}{\ln(-q)^3} \,, 
\end{equation}
%
%
diverges.

\begin{figure}[hbtp]
  \begin{center}
    \scalebox{0.35}{\leavevmode \epsfbox[200 0 400 570]{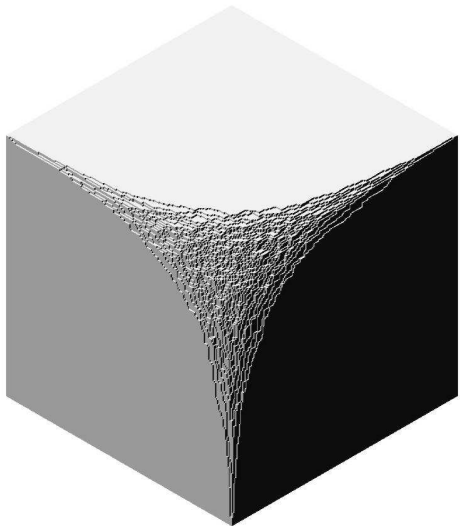}}
      \caption{A random 3D partition of a large number}\label{f_lim}
     \end{center}
\end{figure}

In general, the words ``thermodynamic limit'' have to be taken 
with a grain of salt since $\bw$ is not necessarily a 
positive measure. However, for example in the uniform measure 
case \eqref{CY} it is positive for $-q \in (0,1)$.  
After scaling by $-\ln(-q)$ in every direction, 
a macroscopic limit shape emerges. 
A simulation of the limit shape can be seen in Figure \ref{f_lim}. 

The limit shape dominates the partition function $Z_{DT}$.
The Gromov-Witten partition function $Z_{GW}$ is 
determined by the fluctuations around the limit shape.

\subsubsection{}
The limit shape of a uniformly random 3D partition of
a large number, first determined in \cite{CKen}, is, as it 
turns out, nothing 
but the so-called Ronkin function of the simplest
plane curve
\begin{equation}
  \label{str}
    z+w=1\,,
\end{equation}
see \cite{KOS} for a much more general result. 

Surprisingly (or not ?) the straight line \eqref{str} 
is essentially the Hori-Vafa mirror \cite{HorV} of $\C^3$, see e.g.\
Section 2.5 in \cite{topver}. The mirror
geometry thus can be interpreted as the limit shape
in the localization formula for the original 
counting problem. 

This phenomenon was first observed in \cite{NO} in 
the context of supersymmetric gauge theories on $\R^4$.  
Namely, in \cite{NO} the Seiberg-Witten curve was identified 
with the limit shape in a certain random partition 
ensemble originating from localization 
on the instanton moduli spaces \cite{N}. 
This limit shape interpretation gave a 
a gauge-theoretic derivation of
the Seiberg-Witten prepotential, see \cite{NO}
and also \cite{NY} for a 
different approach. 
Via a physical procedure called \emph{geometric
engineering}, supersymmetric gauge theories correspond to 
Gromov-Witten theory of certain noncompact toric
Calabi-Yau threefolds $X$, see for example \cite{KKV,IK}. 

For toric Calabi-Yau $X$, the random 
surface model can be viewed as a very degenerate
limit of the planar dimer model. There is general
method for finding limit shapes in the dimer 
model, which often gives essentially algebraic 
answers \cite{KO}. In particular, it reproduces the 
Hori-Vafa mirrors of toric Calabi-Yau 3-folds \cite{KOV}. 
It would be extremely interesting to extend the 
$$
\textup{``mirror geometry = limit shape''}
$$
philosophy to a more 
general class of varieties and/or theories.

\subsubsection{}
A natural set of observables to average
against the equivariant vertex measure is 
provided by the characteristic classes of 
the universal sheaf $\cI$, see Section \ref{s_unis}, 
in particular, by the components $\ch_k(\cI)$ of 
its Chern character. The restriction $\ch_k(\pi)$ of $\ch_k(\cI)$ 
to a fixed point $I_\pi\in\Hilb(\C^3;0,\chi)$ 
is determined in terms of the generating 
function \eqref{R_pi} by 
$$
\sum_{k} \alpha^k \, \ch_k(\pi) = 
\frac{R_\pi(e^\alpha)}{R_\varnothing(e^\alpha)} \,.
$$
The algebra generated by $\ch_k(\pi)$ can be 
viewed as the algebra of symmetric polynomials 
in $\pi$; this is a 3-dimensional analog of the 
algebra introduced in \cite{KerO}. 

We have $\ch_1(\pi)=0$, $\ch_2(\pi)=\textup{degree}=0$, and 
$$
\ch_3(\pi) = t_1 t_2 t_3 |\pi|\,,
$$
so from \eqref{DTMc} we get the evaluation 
$$
\langle \ch_3(\pi) \rangle_\bw =
 - (t_1+t_2)(t_1+t_3)(t_2+t_3) \, E_3(-q) \,.
$$
Here $E_{2k+1}$ are the following ``odd weight'' 
analogs of the classical Eisenstein series
\begin{equation}
  \label{E2k+1}
  E_{2k+1}(q) = \sum_n q^n \sum_{d|n} d^{2k} \,, \quad k=1,2,\dots\,.
\end{equation}
One further computes, for example, 
\begin{equation*}
\langle \ch_4(\pi) \rangle_\bw =
 - \frac12 \, (t_1+t_2)(t_1+t_3)(t_2+t_3)
(t_1+t_2+t_3) \,
q\frac{d}{dq} E_3(-q) \,,  
\end{equation*}
and the natural conjecture is that all $\langle \ch_k(\pi) \rangle_\bw$
belong to the differential algebra generated by the 
functions \eqref{E2k+1} and the operator $q\frac{d}{dq}$. A similar 
statement for ordinary 2D partitions and usual even 
weight Eisenstein series was proven in \cite{BO}.

Note, in particular, this conjecture 
implies that the ``thermodynamic'' asymptotics 
of $\langle \ch_k(\pi) \rangle_\bw$ as $q\to -1$ 
is completely determined by the 
first few coefficients of its ``low temperature'' 
$q$-expansion. For a complete
$3$-fold $X$, a similar property is implied by the 
conjectural rationality of the reduced 
partition function $Z'_{DT}$.

\subsubsection{}
Recall that on the Gromov-Witten side, the observables 
$\ch_k(\cI)$ are supposed to correspond to descendent 
invariants. While working out an exact match, especially 
in the equivariant theory, remains an open problem 
(see the discussion in \cite{mnop}), there is one case 
that we understand well. 

Let $X=\Pl\times \C^2$ and let $\beta$ be $d$ times the 
class of $\Pl\times\{0\}$. Let $\C^*$ act on $\C^2$ with 
opposite weights. The $\C^*$-equivariant 
Gromov-Witten theory of $X$ is the Gromov-Witten 
of $\Pl$ with additional insertion of two 
Chern polynomials of the Hodge bundle. Because
of our choice of weights and Mumford's relation, 
these Chern polynomials cancel out, leaving 
us with the Gromov-Witten theory of $\Pl$. 

A complete description of the Gromov-Witten
theory of $\Pl$ was obtained in \cite{OP2,OP3,OP4}. 
In particular, we have the following formula
for disconnected, degree $d$ descendent 
invariants of the point class 
\begin{equation}
  \label{P1}
  \lan \prod \tau_{k_i}(pt) \ran_d^{\Pl} = 
\sum_{|\lambda|=d} \left(\frac{\dim \lambda}{d!}\right)^2 \, 
\prod_i\frac{\bp_{k_i+1}(\lambda)}{(k_i+1)!} \,,
\end{equation}
where the summation is over partitions $\lambda$ of $d$, 
$\dim \lambda$ is the dimension of the corresponding 
representation of the symmetric group, and $\bp_k$ is 
the following polynomial of $\lambda$
\begin{align*}
  \label{bpk}
  \bp_k(\lambda) & \,=\sum_i \left[ (\lambda_i -i + \tfrac12)^k -
(-i + \tfrac12)^k\right] + (1-2^{-k}) \zeta(-k) \\
&\textup{``=''} \sum_i (\lambda_i -i + \tfrac12)^k\,. 
\end{align*}
Here the first line is the $\zeta$-regularization of the 
divergent sum in the second line. The weight function 
in \eqref{P1} is known as the \emph{Plancherel measure} 
on partitions of $d$. 

Sums of the form \eqref{P1} are distinguished 
discrete analogs of 
matrix integrals mentioned at the very beginning 
of the lecture, see e.g.\ the discussion in \cite{uses}. 

\begin{figure}[hbtp]
  \begin{center}
   \scalebox{0.5}{\includegraphics{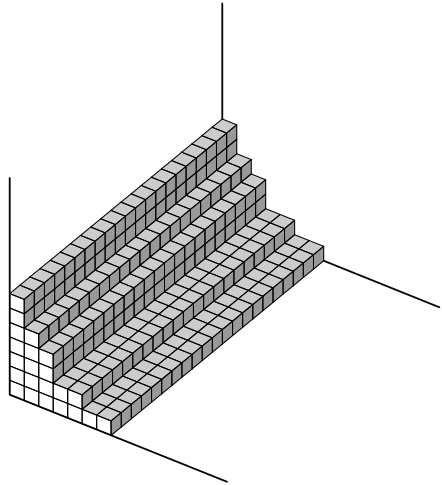}} 
      \caption{A pure edge}\label{f6}
     \end{center}
\end{figure}

What happens on the Donaldson-Thomas side is that with 
our choice of torus weights the contribution of most 
$\bT$-fixed points to the localization formula 
vanishes. The only remaining ones are of the 
form seen in Figure \ref{f6}, they are pure edges, that is, 
cylinders over
an ordinary partition $\lambda$. 

Sure enough, the localization weight of such a pure edge 
in this case specializes to the Plancherel weight of 
its cross-section $\lambda$. Also, the restrictions of $\ch_k(\cI)$ 
to such a fixed point has a simple linear relation to 
the numbers $\bp_k(\lambda)$. 

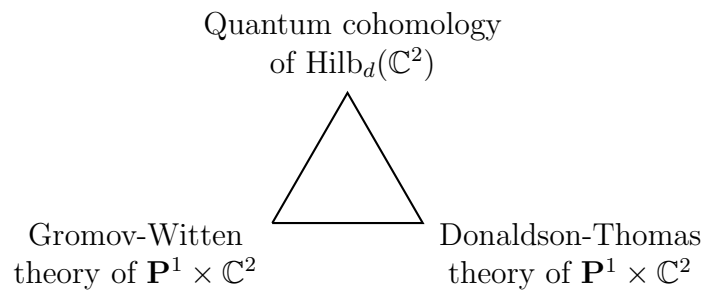
\begin{figure}[hbtp]\psset{unit=0.5 cm}
  \begin{center}
    \begin{pspicture}(-6,-2)(10,6)
    \psline(0,0)(2,3.464)(4,0)(0,0) 
    \rput[rt](0,0){
        \begin{minipage}[t]{3.64 cm}
          \begin{center}
            Gromov-Witten \\ theory of $\Pl \times \C^2$
          \end{center}
        \end{minipage}}
    \rput[lt](4,0){
        \begin{minipage}[t]{3.64 cm}
          \begin{center}
             Donaldson-Thomas\\ theory of $\Pl \times \C^2$
          \end{center}
        \end{minipage}}
    \rput[cb](2,4.7){
        \begin{minipage}[t]{4 cm}
          \begin{center}
           Quantum cohomology \\ of $\Hilb_d(\C^2)$
          \end{center}
        \end{minipage}}
    \end{pspicture}
    \caption{Three points of view on curves in $\Pl \times \C^2$}
\label{f7}
  \end{center}
\end{figure}

It was noticed by several people, in particular in \cite{LQW,LMN}, 
that the sum \eqref{P1} is closely related to localization 
expressions in the classical cohomology of the Hilbert 
scheme of $d$ points in $\C^2$. Perhaps the best 
explanation for this relation is that it is a specialization 
of the triangle of equivalences in 
Figure \ref{f7}, see \cite{OP6}. 

\end{document}